\documentclass[11pt]{article}

\usepackage{amsmath,tabulary,graphicx,times,caption,fancyhdr}
\usepackage[utf8]{inputenc}
\usepackage[T1]{fontenc}
\usepackage{bm}
\usepackage{algorithm}

\usepackage{caption} 
\usepackage{booktabs} 
\usepackage{array} 
\usepackage{threeparttable} 

\usepackage{algorithmic}
\usepackage{natbib}
\usepackage{fancyhdr}
\usepackage{amssymb}
\usepackage{hyperref}
\hypersetup{
    colorlinks=true,
    linkcolor=blue,
    citecolor = blue,
    filecolor=magenta,  
    urlcolor=cyan
    }

\usepackage{url,multirow,morefloats,floatflt,cancel,tfrupee}

\usepackage{colortbl}
\usepackage{xcolor}
\usepackage{pifont}
\usepackage[nointegrals]{wasysym}

\usepackage{authblk}

\title{Ensemble Survival Analysis for Preclinical Cognitive Decline Prediction in Alzheimer's Disease Using Longitudinal Biomarkers}
\author[1]{Dhrubajyoti Ghosh}
\author[2]{Samhita Pal}
\author[3]{Michael Lutz}
\author[1]{Sheng Luo}

\affil[1]{Department of Biostatistics and Bioinformatics, Duke University}
\affil[2]{Department of Statistics, North Carolina State University}
\affil[3]{Department of Neurology, Duke University School of Medicine}

\date{}

\begin{document}

\maketitle

\begin{abstract}
\noindent \textbf{Background:} Predicting the risk of clinical progression from cognitively normal (CN) status to mild cognitive impairment (MCI) or Alzheimer’s disease (AD) is critical for early intervention in Alzheimer’s disease (AD). Traditional survival models often fail to capture complex longitudinal biomarker patterns associated with disease progression.\\
\textbf{Objective:} We propose an ensemble survival analysis framework integrating multiple survival models to improve early prediction of clinical progression in initially cognitively normal individuals.\\
\textbf{Methods:} We analyzed longitudinal biomarker data from the Alzheimer’s Disease Neuroimaging Initiative (ADNI) cohort, including 721 participants, limiting analysis to up to three visits (baseline, 6-month follow-up, 12-month follow-up). Of these, 142 (19.7\%) experienced clinical progression to MCI or AD. Our approach combined penalized Cox regression (LASSO, Elastic Net) with advanced survival models (Random Survival Forest, DeepSurv, XGBoost). Model predictions were aggregated using ensemble averaging and Bayesian Model Averaging (BMA). Predictive performance was assessed using Harrell’s concordance index (C-index) and time-dependent area under the curve (AUC).\\
\textbf{Results:} The ensemble model achieved a peak C-index of 0.907 and an integrated time-dependent AUC of 0.904, outperforming baseline-only models (C-index 0.608). One follow-up visit after baseline significantly improved prediction accuracy (48.1\% C-index, 48.2\% AUC gains), while adding a second follow-up provided only marginal gains (2.1\% C-index, 2.7\% AUC).\\
\textbf{Conclusions:} Our ensemble survival framework effectively integrates diverse survival models and aggregation techniques to enhance early prediction of preclinical AD progression. These findings highlight the importance of leveraging longitudinal biomarker data, particularly one follow-up visit, for accurate risk stratification and personalized intervention strategies.
\end{abstract}

\section{Introduction}

Alzheimer’s disease (AD), the leading cause of dementia, affects approximately 6.9 million Americans aged 65 and older in the United States, with projections suggesting a doubling by 2060 absent effective interventions \cite{2024ADJ_facts_figures}. Defined by amyloid-$\beta$ (A$\beta$) plaques, tau neurofibrillary tangles, and neuronal loss, AD progresses from preclinical stages to cognitive decline \cite{blennow2010cerebrospinal}. Mild cognitive impairment (MCI) often precedes AD, with many cognitively normal (CN) individuals advancing to MCI or AD within years \cite{Petersen1999AN}. Early risk detection enables interventions like lifestyle modifications, pharmacological therapies, and clinical trials \cite{sperling2014evolution}, yet predicting this transition remains complex due to progression heterogeneity and intricate biomarker interactions \cite{jack2010hypothetical}.

Conventional AD prediction relies on cross-sectional neuropsychological assessments, such as the Mini-Mental State Examination (MMSE) and Clinical Dementia Rating (CDR), which often fail to detect preclinical shifts \cite{Albert2011ADJ, Petersen1999AN}. Longitudinal biomarkers, including neuroimaging, cerebrospinal fluid (CSF), and repeated cognitive measures derived from these tests (e.g., MMSE scores over time), provide dynamic insights, increasing accuracy \cite{Jack2013Lancet-Neurology, Parent2023Aging-Brain}. Research shows biomarker trajectories improve risk stratification \cite{Li2017JAD}, with joint longitudinal-survival models surpassing baseline-only methods \cite{Li2018ADJ, li2019dynamic}. However, focus on MCI-to-AD progression often dominates attention on the subtler CN-to-MCI/AD transition, insufficiently researched due to its gradual onset and variable presentation \cite{chen2017progression, cui2011identification, blacker2007neuropsychological}. Single-model approaches, even with multimodal data, face challenges in capturing AD's complexity, as preclinical changes involve nonlinear relationships across genetic, imaging, and clinical domains \cite{Jack2013Lancet-Neurology, yao2022ensemble}, requiring advanced ensemble survival strategies.

Ensemble learning, combining multiple models for robustness, excels in AD classification \cite{syed2020ensemble, javeed2022intelligent, ghali2020advanced, wang2018early}, yet its survival analysis use is limited. Hothorn et al.'s survival ensembles \cite{hothorn2006survival} enhanced predictions in prostate \cite{polsterl2016heterogeneous} and skin cancer \cite{abbasi2023optimizing}, and Yao et al.'s longitudinal covariate ensemble improved accuracy \cite{yao2022ensemble}. Despite these advances, CN-to-MCI/AD progression prediction via ensembles remains inadequately investigated. We propose an ensemble survival framework merging penalized Cox regression with the Least Absolute Shrinkage and Selection Operator (LASSO), Elastic Net, Random Survival Forest, DeepSurv, and XGBoost, leveraging multimodal longitudinal biomarkers and aggregation techniques (ensemble averaging, Bayesian Model Averaging) to predict preclinical AD progression accurately. This approach integrates linear and nonlinear modeling to address AD's heterogeneity. It uses longitudinal data from the Alzheimer’s Disease Neuroimaging Initiative (ADNI) cohort of 721 cognitively normal individuals to detect subtle preclinical shifts in those progressing to MCI or AD. It aims to enhance early risk stratification, providing a robust predictive tool validated through rigorous metrics, potentially guiding timely interventions in AD's preclinical phase.

This article is structured as follows: Section~\ref{sec:study} details the ADNI cohort and data processing; Section~\ref{sec:methods} outlines methodology; Section~\ref{sec:results} presents results; Section~\ref{sec:conclusion} discusses implications, limitations, and future directions.

\section{Cohort and Data Processing} \label{sec:study}

\subsection{Study Population}

The ADNI study is a longitudinal study assessing serial neuroimaging, clinical, and neuropsychological measures to track AD progression (\href{http://www.adni-info.org}{www.adni-info.org}). ADNI’s primary goal is to test whether serial magnetic resonance imaging (MRI), positron emission tomography (PET), other biological markers, and clinical and neuropsychological assessments can be combined to measure the progression of MCI and early AD. Using ADNI data, the current study predicts time to clinical progression from CN to MCI or AD over the full follow-up period, targeting preclinical AD. CN, MCI, and AD were defined per ADNI criteria, with CN requiring MMSE 24–30 and CDR 0, MCI per Petersen et al. \cite{Petersen1999AN} with CDR 0.5, and AD per NINCDS/ADRDA probable AD criteria \cite{McKhann1984Neurology}, requiring memory impairment, deficits in at least one other cognitive domain (e.g., MMSE 20–26 for global cognition), and functional impairment (e.g., Functional Activities Questionnaire, FAQ >9, CDR 0.5–1 for mild to moderate severity). We analyzed longitudinal biomarker data from 721 CN participants in the ADNI study, of whom 142 (19.7\%) progressed (104 to MCI, 38 to AD), requiring at least two follow-up visits for longitudinal analysis. Analysis utilized data from up to three visits per participant: baseline, a 6-month follow-up (Visit 1), and a 12-month follow-up (Visit 2), based on ADNI-1’s schedule of assessments at baseline, 6, 12, 18, 24, and 36 months, with annual follow-ups in ADNI-2. Time between visits was calculated as the difference in months between recorded visit dates in ADNI’s dataset, yielding intervals of approximately 6 months (baseline to Visit 1, mean 6.79 months, SD 3.63 months, with variability reflecting real-world scheduling) and 12 months (baseline to Visit 2, mean 13.32 months, SD 7.25 months, with variability reflecting real-world scheduling), reflecting typical ADNI visit adherence. Characteristics are in Table~\ref{tab:char-table}.

\subsection{Biomarker Data}

Predictors included baseline variables, such as demographic factors (age, gender, race, ethnicity, marital status, education, APOE4 genotype) linked to AD risk \cite{jack2010hypothetical} and CSF biomarkers (A$\beta$, total tau, phosphorylated tau [p-tau]) primarily measured at baseline. Longitudinal variables included neuroimaging biomarkers (hippocampal, intracranial, ventricle, mid-temporal, fusiform gyrus, entorhinal cortex, whole brain volumes), PET biomarkers (fluorodeoxyglucose [FDG], Pittsburgh Compound B [PIB], AV45), and cognitive assessments (MMSE, Alzheimer's Disease Assessment Scale [ADAS13], Montreal Cognitive Assessment [MOCA], FAQ, Rey Auditory Verbal Learning Test [RAVLT] components), all measured longitudinally and reflecting AD pathology, amyloid/tau accumulation, metabolism/amyloid burden, and cognitive function, respectively. To detect preclinical cognitive decline, we examined baseline measures and longitudinal biomarker trajectories, calculating rates of change as differences between consecutive visits divided by time intervals (e.g., $\Delta_{0 \rightarrow 1} = (x_1 - x_0) / (t_1 - t_0)$).

\subsection{Data Preprocessing}

Missing data were imputed using Multiple Imputation by Chained Equations (MICE) \cite{White2011SIM_MICE}, generating 20 datasets over 1000 iterations, pooled via Rubin's Rules (Section~\ref{sec:methods-imputation}). Continuous variables were standardized (zero mean, unit variance), categorical variables one-hot encoded, and penalized regression selected predictive features (Section~\ref{sec:methods-feature}).

\subsection{Longitudinal Modeling}

Three scenarios evaluated longitudinal biomarker impacts on preclinical progression: baseline features only, baseline plus visit 1 changes ($\Delta_{0 \rightarrow 1}$), and baseline features plus visit 1 changes ($\Delta_{0 \rightarrow 1}$) and visit 1 and 2 changes ($\Delta_{1 \rightarrow 2} = (x_2 - x_1) / (t_2 - t_1)$), using data from up to three visits (baseline, visit 1, visit 2). The analysis was limited to three visits to focus on early predictive patterns, with further details on performance trends provided in Section~\ref{sec:results-performance}.

\subsection{Data Splitting and Validation}

The dataset was split into 80\% training and 20\% testing sets, with 5-fold cross-validation ensuring robustness. Training fitted ensemble models (Section~\ref{sec:methods-survival}), while testing validated performance (Section~\ref{sec:results}), leveraging longitudinal trajectories to enhance early risk stratification.

\begin{table}[htbp]
\centering
\rowcolors{2}{white}{gray!10}
\resizebox{0.7\textwidth}{!}{%
\begin{tabular}{lll}
\hline
Variable & Did not Progress (N=579) & Clinical Progression (N=142) \\ \hline
Age & 72.52 (6.17) & 74.94 (5.56) \\
Education & 16.59 (2.51) & 16.22 (2.54) \\
Gender (Female) & 57.68\% & 47.18\% \\
\multicolumn{1}{l}{Race:} & & \\
\ \ Indian/Alaskan & 0.34\% & - \\
\ \ Asian & 1.90\% & 2.11\% \\
\ \ Black & 5.53\% & 6.34\% \\
\ \ More than 1 & 1.73\% & 1.41\% \\
\ \ White & 90.50\% & 90.14\% \\
\multicolumn{1}{l}{Ethnicity:} & & \\
\ \ Hispanic/Latino & 3.97\% & 1.41\% \\
\ \ Not Hispanic/Latino & 95.34\% & 98.59\% \\
\ \ Unknown & 0.69\% & - \\
\multicolumn{1}{l}{Marital Status:} & & \\
\ \ Divorced & 11.23\% & 9.12\% \\
\ \ Married & 70.29\% & 73.95\% \\
\ \ Never married & 5.18\% & 4.81\% \\
\ \ Widowed & 13.13\% & 11.54\% \\
\ \ Unknown & 0.17\% & - \\
FAQ & 0.15 (0.60) & 0.36 (1.17) \\
MMSE & 29.15 (1.09) & 28.95 (1.14) \\
FDG & 1.27 (0.08) & 1.25 (0.11) \\
PIB & 1.60 (0.00) & 1.60 (0.00) \\
AV45 & 1.11 (0.12) & 1.15 (0.14) \\
Hippocampus volume & 7537.19 (852.69) & 7138.83 (759.71) \\
ICV & 1490394.11 (157968.53) & 1522098.00 (160225.30) \\
Mid-temporal volume & 20793.04 (2661.18) & 20109.59 (2527.71) \\
Fusiform gyrus volume & 18361.99 (2350.49) & 17732.87 (2142.32) \\
Ventricle volume & 32403.32 (16931.73) & 38655.70 (20581.86) \\
Entorhinal cortex volume & 4001.03 (627.27) & 3745.27 (658.49) \\
Whole brain volume & 1044926.66 (104565.72) & 1028377.00 (98540.06) \\
APOE4 & 0.31 (0.50) & 0.40 (0.56) \\
A$\beta$ & 1033.20 (219.60) & 968.40 (267.36) \\
Tau & 242.49 (58.06) & 255.03 (74.25) \\
p-Tau & 22.36 (5.90) & 23.81 (7.79) \\
RAVLT learning & 6.27 (2.27) & 5.38 (2.45) \\
RAVLT forgetting & 3.62 (2.84) & 4.10 (2.80) \\
RAVLT immediate & 46.75 (9.82) & 41.17 (8.82) \\
LDELTOTAL & 13.41 (3.34) & 12.17 (3.28) \\
mPACCdigit & 0.29 (2.61) & -1.03 (2.82) \\
mPACCtrailsB & 0.33 (2.44) & -1.05 (2.57) \\
ADAS13 & 8.36 (4.14) & 10.69 (4.68) \\
MOCA & 26.13 (2.09) & 24.94 (1.76) \\ \hline
\end{tabular}%
}
\caption{Demographic and clinical characteristics at baseline, grouped by progression status. Data are mean (SD) or \% unless specified. Clinical Progression includes progression to Mild Cognitive Impairment (MCI, N=104) or Alzheimer's Disease (AD, N=38). ADAS13, Alzheimer's Disease Assessment Scale; MOCA, Montreal Cognitive Assessment; FAQ, Functional Activities Questionnaire; MMSE, Mini-Mental State Examination; FDG, Fluorodeoxyglucose; PIB, Pittsburgh Compound B; AV45, Florbetapir F18; ICV, Intracranial Volume; APOE4, Apolipoprotein E epsilon 4 genotype; A$\beta$, Amyloid-beta; p-Tau, Phosphorylated Tau; RAVLT, Rey Auditory Verbal Learning Test; LDELTOTAL, Logical Memory Delayed Recall Total Score; mPACCdigit, Modified Preclinical Alzheimer Cognitive Composite - Digit Symbol Substitution; mPACCtrailsB, Modified Preclinical Alzheimer Cognitive Composite - Trails B Test.}
\label{tab:char-table}
\end{table}

\section{Ensemble Survival Analysis for Predicting Clinical Progression in AD} \label{sec:methods}

\subsection{Overview}

This study presents an ensemble survival analysis framework to predict the time to clinical progression in AD from CN to MCI or AD, emphasizing preclinical AD as introduced in Sections 1 and 2. The framework integrates feature selection with penalized regression and survival modeling via an ensemble of advanced techniques. Feature selection utilized Cox LASSO \cite{tibshirani1997lasso} and Cox Elastic Net \cite{wu2012elastic}, extending traditional Cox proportional hazards (Cox PH) models with regularization penalties to handle high-dimensional data. Survival modeling employed Random Survival Forest (RSF) \cite{ishwaran2008random}, Deep Survival Analysis (DeepSurv) \cite{katzman2018deepsurv}, and Extreme Gradient Boosting (XGBoost) \cite{chen2016xgboost}, selected for their capacity to model complex relationships in longitudinal biomarker trajectories from the ADNI dataset, comprising 721 CN participants in the ADNI cohort, using longitudinal data from up to three visits per participant (baseline, visit 1, and visit 2) as detailed in Section~\ref{sec:study}. Standalone Cox PH modeling was excluded due to overfitting risks given the dataset's dimensionality relative to sample size. Risk predictions from individual models were aggregated using Ensemble Averaging (EA) and Bayesian Model Averaging (BMA) to improve accuracy and robustness, with detailed steps outlined below.

\subsection{Regularized Feature Selection} \label{sec:methods-feature}

To identify predictors most relevant to survival modeling, we applied Cox PH models to the multimodal ADNI dataset\footnote{Data used in the preparation of this article were obtained from the Alzheimer’s Disease Neuroimaging Initiative (ADNI) database (\href{adni.loni.usc.edu}{adni.loni.usc.edu}). The ADNI was launched in 2003 as a public-private partnership, led by Principal Investigator Michael W. Weiner, MD. The primary goal of ADNI has been to test whether serial magnetic resonance imaging
(MRI), positron emission tomography (PET), other biological markers, and clinical and neuropsychological assessment can be combined to measure the progression of mild
cognitive impairment (MCI) and early Alzheimer’s disease (AD). For up-to-date information,
see \href{www.adni-info.org.}{www.adni-info.org.}}, defining time to clinical progression as the outcome and assigning maximum follow-up time to censored cases. With 721 participants and extensive longitudinal biomarkers (Section~\ref{sec:study}), we addressed multicollinearity and overfitting using penalized regression. Cox LASSO \cite{tibshirani1997lasso} applies an $\ell_1$ penalty, shrinking less predictive coefficients to zero to promote sparsity, which is advantageous in high-dimensional survival datasets where many covariates may be redundant or weakly associated with the hazard. Cox Elastic Net \cite{wu2012elastic} uses $\ell_1$ and $\ell_2$ penalties to balance sparsity and stability. It groups correlated predictors, like neuroimaging or CSF measures, to improve interpretability and predictive power for AD's complex data. These methods selected key features, with results detailed in Section~\ref{sec:results}.

\subsection{Survival Models for Clinical Progression} \label{sec:methods-survival}

Survival analysis focused on time to first clinical progression, including only baseline CN participants and censoring non-progressors at their last visit in the ADNI longitudinal record, while predictors were assessed over three visits (baseline, visit 1, and visit 2) as described in Section~\ref{sec:study}. We assessed three scenarios to evaluate longitudinal biomarker trajectories' impact: baseline features only, baseline plus visit 1 changes ($\Delta_{0 \rightarrow 1}$), and baseline plus visit 1 and 2 changes ($\Delta_{1 \rightarrow 2} = (x_2 - x_1) / (t_2 - t_1)$), building on data from Section~\ref{sec:study}. The ensemble included RSF \cite{ishwaran2008random}, which extends random forests to survival by constructing decision trees estimating survival probabilities without hazard assumptions, ideal for capturing complex interactions in high-dimensional data. DeepSurv \cite{katzman2018deepsurv} leverages neural networks with ReLU activation and dropout to model nonlinear covariate-survival relationships, enhancing generalization. XGBoost \cite{chen2016xgboost} uses gradient boosting to refine decision trees sequentially, optimizing survival predictions with regularization for robustness. See Appendix~\ref{appendix} for mathematical details.

\subsection{Ensemble Risk Score Aggregation} \label{sec:methods-ensemble}

Each model produced risk scores reflecting progression likelihood, varying due to their unique strengths. We applied EA and BMA to aggregate these scores. EA computes $\hat{\beta}_{EA,i} = \frac{1}{3} \sum_{m=1}^3 \hat{\beta}_{m,i}$, where $M=3$ and $\hat{\beta}_{m,i}$ is the $m$-th model's score for individual $i$, averaging equally to reduce variance. BMA weights scores by posterior probabilities, $\hat{\beta}_{BMA,i} = \sum_{m=1}^M w_m \hat{\beta}_{m,i}$, with $\sum_{m=1}^M w_m = 1$, reflecting model uncertainty via data likelihood and priors, yielding refined accuracy (Section~\ref{sec:results}). 

\subsection{Evaluation Metrics for Predictive Performance} \label{sec:methods-evaluation}

We evaluated performance using Harrell's C-index \cite{harrell1982evaluating} and time-dependent Area Under the Curve (AUC) \cite{heagerty2000time} across 20 imputed datasets (Section~\ref{sec:methods-imputation}). The C-index, $C = P(\hat{S}(T_i|x_i) < \hat{S}(T_j|x_j) | T_i < T_j, \delta_i = 1)$, where $\hat{S}(t|X)$ is the survival function and $\delta_i$ indicates event (1) or censoring (0), quantifies ranking accuracy from 0.5 (random) to 1.0 (perfect), leveraging its nonparametric nature for survival analysis. Time-dependent AUC, $\text{AUC}(t) = P(\hat{h}_i > \hat{h}_j | T_i < T_j, T_i < t)$, was computed at 100 evenly spaced time points across the observed range of survival times (\( T_{\min} \) to \( T_{\max} \)), using inverse probability of censoring weighting (IPCW) to adjust for censoring bias, evaluating discrimination between individuals experiencing events by time \( t \) and those at risk. Integrated AUC (iAUC), $\bar{\text{AUC}} = \frac{1}{T_{\max} - T_{\min}} \int_{T_{\min}}^{T_{\max}} \text{AUC}(t) dt$, was calculated via numerical integration using the trapezoidal rule, providing a comprehensive discrimination measure for longitudinal data. These metrics, detailed with results in Tables~\ref{tab:longitudinal_penalty_C-index} and \ref{tab:longitudinal_performance_AUC}, ensure robust assessment of ranking and calibration.

\subsection{Multiple Imputation and Rubin's Method} \label{sec:methods-imputation}

Missing data, common in AD longitudinal studies, were handled using MICE \cite{White2011SIM_MICE}, assuming Missing at Random (MAR) or Missing Completely at Random (MCAR) mechanisms. MAR implies missingness depends on observed data, while MCAR assumes randomness unrelated to any data. MICE generated 20 datasets over 1000 iterations using Bayesian linear regression \cite{minka2000bayesian}, modeling each variable $X_j$ with missing values $X_{\text{mis}}$ as $X_j^{(t+1)} = f_j(X_{-j} | \theta_j) + \epsilon_j$, where $f_j$ is the regression function, $X_{-j}$ are other variables, $\theta_j$ are parameters, and $\epsilon_j \sim \mathcal{N}(0, \sigma_j^2)$ adds uncertainty, iterating until convergence. Rubin's Rules pooled estimates as $\bar{\theta} = \frac{1}{M} \sum_{m=1}^M \hat{\theta}_m$, with total variance $T = W + (1 + \frac{1}{M})B$, where within-variance $W = \frac{1}{M} \sum_{m=1}^M \text{Var}(\hat{\theta}_m)$, between-variance $B = \frac{1}{M-1} \sum_{m=1}^M (\hat{\theta}_m - \bar{\theta})^2$, and $M=20$. Confidence intervals used Student's $t$-distribution, with degrees of freedom $\nu = \big({(M - 1) \big( 1 + \frac{W}{(1 + 1/M) B} \big)^2}\big)/\big({1 + \frac{W}{(1 + 1/M) B}}\big)$, applied to C-index and iAUC (Section~\ref{sec:results}).

\section{Results} \label{sec:results}

\subsection{Feature Selection and Longitudinal Biomarker Dynamics} \label{sec:results-feature-selection}

We identified key predictors for survival modeling using Cox LASSO and Cox Elastic Net, applied to longitudinal biomarker data from 721 CN participants in the ADNI cohort, using data from up to three visits per participant (baseline, visit 1, and visit 2) as detailed in Section~\ref{sec:study}. Table~\ref{tab:var_sel} summarizes features at each longitudinal stage. Baseline included 51 variables, increasing to 75 with visit 1 ($\Delta_{0 \rightarrow 1}$) and 99 with visit 2 ($\Delta_{1 \rightarrow 2}$), reflecting added rate-of-change variables. Across 20 imputed datasets (Section~\ref{sec:methods-imputation}), feature counts varied slightly due to missing data patterns, with the numbers in Table~\ref{tab:var_sel} representing averages pooled via Rubin's Rules.

\begin{table}[htbp]
\centering
\caption{Number of features included at each longitudinal stage (baseline, two visits, and three visits) and subsets selected by each penalization method.}
\label{tab:var_sel}
\resizebox{0.6\textwidth}{!}{%
\begin{tabular}{cccc}
\hline
Longitudinal Stage & Total Features & Selected by Cox LASSO & Selected by Cox Elastic Net \\ \hline
Baseline & 51 & 35 & 40 \\
2 Visits & 75 & 45 & 53 \\
3 Visits & 99 & 54 & 59 \\ \hline
\end{tabular}%
}
\end{table}

Baseline features selected by Elastic Net commonly included demographics (e.g., age, education), biomarkers (e.g., APOE4, FDG, AV45, p-tau), and cognitive scores (e.g., RAVLT learning, MOCA). Visit 1 added rate-of-change variables ($\Delta_{0 \rightarrow 1}$) like FAQ, MMSE, and cognitive changes, while visit 2 incorporated $\Delta_{1 \rightarrow 2}$ for biomarkers (e.g., p-tau) and neuroimaging volumes (e.g., ICV, hippocampus). These predictors, consistently selected across 20 imputed datasets, reflect AD progression dynamics, with detailed importance in Section~\ref{sec:results-predictors} and Figure~\ref{fig:ea-imp}. We focused on Elastic Net-selected features due to their superior predictive performance (Section~\ref{sec:results-performance}) and overlap with LASSO selections, avoiding redundancy in a high-dimensional dataset where listing all features would be cumbersome.

\subsection{Model Performance} \label{sec:results-performance}

We evaluated ensemble survival models' predictive performance using C-index and time-dependent AUC, assessing longitudinal impacts on preclinical AD progression risk, with events (conversion from CN to MCI/AD) occurring over the full follow-up period, while predictors were limited to data from baseline, the 6-month follow-up (Visit 1), and the 12-month follow-up (Visit 2). Three scenarios were tested: baseline features, baseline plus visit 1 changes, and baseline plus visits 1 and 2 changes, using 5-fold cross-validation across 20 imputed datasets (Section~\ref{sec:methods-evaluation}).

\begin{table}[htbp]
\centering
\caption{Concordance index (C-index) for different longitudinal data settings, penalization methods, and ensemble aggregation techniques.}
\label{tab:longitudinal_penalty_C-index}
\resizebox{\textwidth}{!}{%
\begin{tabular}{ccccccc}
\hline 
\begin{tabular}[c]{@{}c@{}}Longitudinal\\ Data\end{tabular} & Penalty 
& RSF & DeepSurv & XGBoost & EA & BMA \\ \midrule 
Baseline & Cox LASSO & 0.512 (0.511, 0.513) & 0.591 (0.569, 0.612) & 0.580 (0.580, 0.580) & 0.601 (0.587, 0.613) & 0.601 (0.587, 0.614) \\
         & Cox Elastic & 0.503 (0.503, 0.503) & 0.595 (0.562, 0.629) & 0.586 (0.586, 0.586) & 0.607 (0.586, 0.625) & 0.608 (0.588, 0.627) \\ \midrule
2 Visits & Cox LASSO & 0.645 (0.645, 0.645) & 0.792 (0.776, 0.809) & 0.878 (0.878, 0.878) & 0.880 (0.874, 0.887) & 0.882 (0.876, 0.887) \\
         & Cox Elastic & 0.648 (0.648, 0.648) & 0.796 (0.771, 0.821) & 0.881 (0.881, 0.881) & 0.884 (0.874, 0.893) & 0.885 (0.876, 0.895) \\ \midrule
3 Visits & Cox LASSO & 0.689 (0.689, 0.689) & 0.819 (0.807, 0.831) & 0.906 (0.906, 0.906) & 0.904 (0.893, 0.914) & 0.907 (0.897, 0.917) \\
         & Cox Elastic & 0.701 (0.701, 0.701) & 0.809 (0.789, 0.823) & 0.902 (0.902, 0.902) & 0.902 (0.891, 0.912) & 0.904 (0.894, 0.914) \\ \bottomrule 
\end{tabular}%
}
\begin{flushleft}
\footnotesize{
RSF = Random Survival Forest, DeepSurv = Deep Neural Network for Survival Analysis, XGBoost = Gradient Boosting Machine, EA = Ensemble Averaging, BMA = Bayesian Model Averaging. Values are mean C-indices across 5-fold cross-validation, with 95\% confidence intervals from Rubin's Rules for multiple imputed datasets. Bolded values indicate the highest C-index per longitudinal setting.
}
\end{flushleft}
\end{table}

\begin{table}[htbp]
\centering
\caption{Time-dependent AUC for different longitudinal data settings, penalization methods, and ensemble aggregation techniques.}
\label{tab:longitudinal_performance_AUC}
\resizebox{\textwidth}{!}{%
\begin{tabular}{ccccccc}
\hline 
\begin{tabular}[c]{@{}c@{}}Longitudinal\\ Information\end{tabular} & Penalty 
& RSF & DeepSurv & XGBoost & EA & BMA \\ \midrule 
Baseline & Cox LASSO & 0.497 (0.497, 0.497) & 0.564 (0.537, 0.591) & 0.587 (0.587, 0.587) & 0.601 (0.584, 0.617) & 0.601 (0.584, 0.617) \\
         & Cox Elastic & 0.516 (0.516, 0.516) & 0.569 (0.525, 0.614) & 0.592 (0.592, 0.592) & 0.606 (0.584, 0.627) & 0.606 (0.584, 0.628) \\ \midrule
2 Visits & Cox LASSO & 0.641 (0.641, 0.641) & 0.787 (0.772, 0.802) & 0.897 (0.897, 0.897) & 0.897 (0.892, 0.902) & 0.898 (0.893, 0.903) \\
         & Cox Elastic & 0.667 (0.667, 0.667) & 0.797 (0.770, 0.825) & 0.897 (0.897, 0.897) & 0.897 (0.887, 0.907) & 0.898 (0.889, 0.907) \\ \midrule
3 Visits & Cox LASSO & 0.730 (0.730, 0.730) & 0.835 (0.822, 0.848) & 0.917 (0.917, 0.917) & 0.920 (0.912, 0.928) & 0.921 (0.914, 0.928) \\
         & Cox Elastic & 0.719 (0.719, 0.719) & 0.823 (0.799, 0.845) & 0.926 (0.926, 0.926) & 0.921 (0.913, 0.928) & 0.922 (0.915, 0.929) \\ \bottomrule 
\end{tabular}%
}
\begin{flushleft}
\footnotesize{
RSF = Random Survival Forest, DeepSurv = Deep Neural Network for Survival Analysis, XGBoost = Gradient Boosting Machine, EA = Ensemble Averaging, BMA = Bayesian Model Averaging. Values are mean time-dependent AUCs across 5-fold cross-validation, with 95\% confidence intervals from Rubin's Rules for multiple imputed datasets. Bolded values indicate the highest AUC per longitudinal setting.
}
\end{flushleft}
\end{table}

At baseline, DeepSurv outperformed XGBoost and RSF, with C-indices ranging from 0.591–0.595 (Cox LASSO and Elastic) for DeepSurv, 0.580–0.586 for XGBoost, and 0.503–0.512 for RSF, reflecting baseline biomarker informativeness. With longitudinal data, XGBoost emerged as the top performer, surpassing RSF and DeepSurv: visit 1 C-indices ranged from 0.878–0.881 for XGBoost, 0.645–0.648 for RSF, and 0.792–0.796 for DeepSurv, while visit 2 reached 0.902–0.906, 0.689–0.701, and 0.809–0.819, respectively, highlighting the value of repeated measurements. Time-dependent AUCs mirrored these trends, improving from 0.606 (baseline) to 0.922 (visit 2) with Elastic Net + BMA, indicating enhanced discrimination over time. Ensemble models (Elastic Net + BMA) consistently excelled, achieving C-indices of 0.608 (baseline), 0.885 (visit 1), and 0.904 (visit 2), and time-dependent AUCs of 0.606, 0.898, and 0.922, leveraging RSF, DeepSurv, and XGBoost to reduce bias and enhance robustness. This represents a significant improvement with one follow-up visit after baseline (48.1\% C-index, 48.2\% AUC gains), particularly at the 6-month follow-up (Visit 1), enabling early risk stratification in clinical practice with minimal monitoring. Conversely, the marginal gains at the 12-month follow-up (Visit 2) (2.1\% C-index, 2.7\% AUC) indicate that extending follow-up beyond 6 months may not substantially improve prediction accuracy, a key finding that supports less frequent long-term monitoring for stable CN individuals. Age subgroup analysis revealed that predictive accuracy decreases with age, with younger individuals (61–70 years) achieving the highest performance, followed by the 71–80 group, and older individuals (81–90 years) showing the lowest, suggesting age-related heterogeneity influences biomarker informativeness. See Supplementary Material for detailed results and performance metrics by age group.

We recommend using baseline and visit 1 data (6-month follow-up) for early risk stratification, balancing accuracy (48.1\% C-index, 48.2\% AUC gains from baseline to visit 1) and clinical feasibility, minimizing unnecessary testing while ensuring timely interventions for preclinical AD.

\subsection{Key Predictors of Clinical Progression} \label{sec:results-predictors}

\begin{figure}[htbp]
    \centering
    \includegraphics[width=\textwidth]{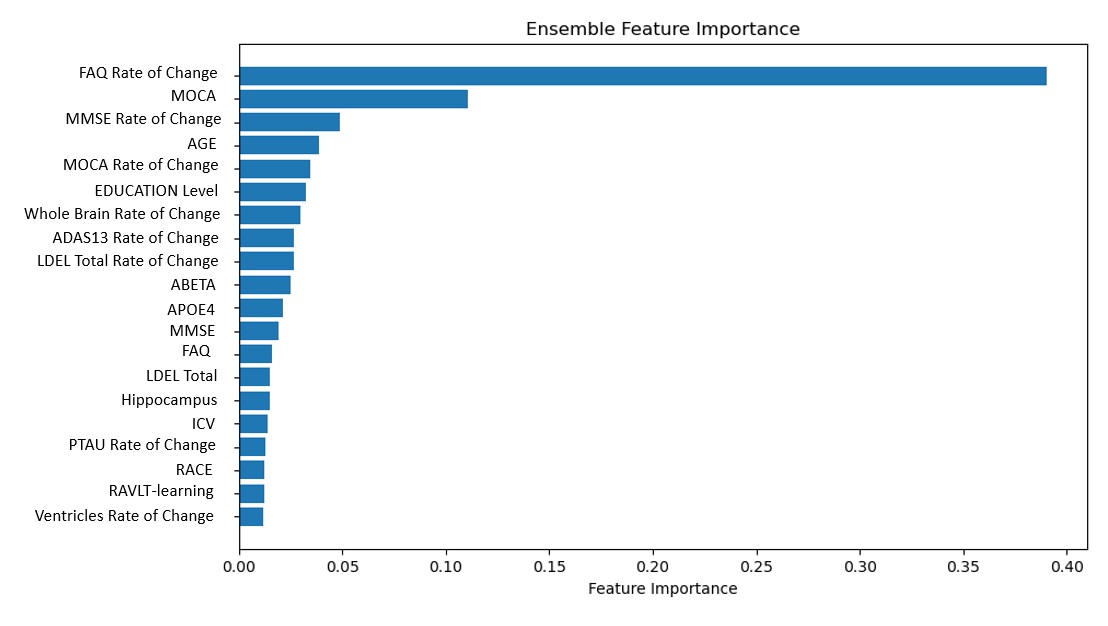}
    \includegraphics[width=\textwidth]{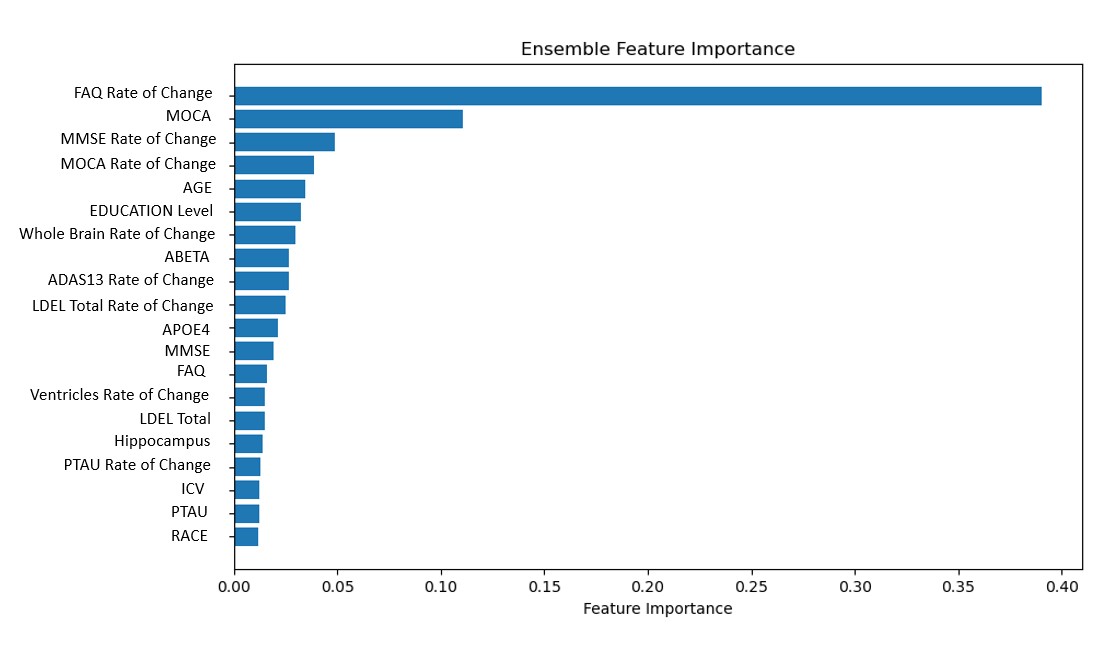}
    \caption{Feature Importance of Ensemble Survival Models Using Different Aggregation Methods. The top panel shows feature importance from ensemble averaging (EA), while the bottom panel shows Bayesian model averaging (BMA). Abbreviations: FAQ, Functional Activities Questionnaire; MOCA, Montreal Cognitive Assessment; MMSE, Mini-Mental State Examination; ADAS13, Alzheimer's Disease Assessment Scale-Cognitive Subscale (13-item version); LDELTOTAL, Logical Memory Delayed Recall Total Score; A$\beta$, Amyloid-beta; APOE4, Apolipoprotein E epsilon 4 genotype; p-tau, Phosphorylated Tau; ICV, Intracranial Volume; RAVLT-learning, Rey Auditory Verbal Learning Test - Learning Score.}
    \label{fig:ea-imp}
\end{figure}

We examined influential predictors of preclinical AD progression using ensemble survival models, focusing on Elastic Net-selected features with visit 1 data, aggregated across 20 imputed datasets via Rubin's method (Section~\ref{sec:methods-imputation}). Figure~\ref{fig:ea-imp} presents feature importance plots, with the top panel for EA and bottom for BMA, reflecting longitudinal biomarker dynamics.

The rate of change in the FAQ was the strongest predictor in both EA and BMA, highlighting its sensitivity to functional decline. Other top predictors included rates of change in MOCA, MMSE, ADAS13, and LDELTOTAL, alongside baseline MOCA and LDELTOTAL, underscoring cognitive decline's role. Demographic factors (age, education) and biomarkers (A$\beta$, APOE4, p-tau rate of change) were critical, with p-tau changes reflecting neurodegeneration as reported in prior studies \cite{blennow2010cerebrospinal}. Neuroimaging features (hippocampal, ventricular, whole-brain volumes, ICV) reflected structural atrophy, and race suggested demographic influences, all enhancing early risk stratification.

\section{Conclusion} \label{sec:conclusion}

\subsection{Summary of Findings}

This study demonstrates that longitudinal data, particularly from visit 1 (6-month follow-up), significantly enhances predictive accuracy for preclinical AD progression. Our ensemble survival models, combining RSF, DeepSurv, and XGBoost with Elastic Net and BMA, achieved a 48.1\% C-index increase (0.608 to 0.885) and 48.2\% time-dependent AUC gain (0.606 to 0.898) with visit 1 data (Section~\ref{sec:results-performance}) over the full follow-up period. The incremental improvement from visit 1 to visit 2 (12-month follow-up) data (2.1\% C-index, 2.7\% AUC) indicates diminishing returns with additional longitudinal data. These findings highlight longitudinal biomarker trajectories' critical role, with Elastic Net and BMA outperforming for robust, generalizable predictions of clinical progression.

\subsection{Clinical Implications}

Enhanced prediction accuracy enables early detection and risk stratification in AD, allowing clinicians to target high-risk CN individuals for timely interventions before cognitive decline, as shown in Section~\ref{sec:results-predictors}. Adaptive monitoring, informed by age, APOE4, cognitive scores, and biomarker changes, optimizes preventive strategies like lifestyle adjustments, therapies, and drugs, improving outcomes. Prioritizing visit 1 data (6-month follow-up) minimizes unnecessary testing while ensuring effective monitoring, addressing preclinical AD's subtle progression (Section~\ref{sec:study}).

\subsection{Implications for Clinical Trials}

Our framework improves AD trial design by predicting time-to-conversion accurately, enabling targeted patient selection to potentially reduce cohort heterogeneity and enhance statistical power based on early risk stratification (Section~\ref{sec:results-performance}). Focusing on the 6-month follow-up, which provides substantial predictive accuracy and optimizes early data collection, reduces costs and timelines while supporting adaptive designs for real-time adjustments, refined criteria, and resource reallocation, accelerating disease-modifying therapy development.

\subsection{Future Directions}

Future research should validate our models externally across populations, integrating multi-modal biomarkers (e.g., genetics, blood, neuroimaging) to enhance predictions. Age-stratified modeling, as shown in Supplementary Material, could further refine predictions, particularly for older individuals with heterogeneous progression. Deep learning survival models could automate feature selection and capture complex patterns, though computational and interpretability challenges need resolution. Incorporating real-world EHRs could enable routine early detection, requiring workflow integration. Studies should assess these models' impact on treatment decisions, outcomes, and policies for preclinical AD management, potentially leveraging diverse datasets like NACC, part of the ADRCs network, to enhance generalizability across populations.

\section*{Acknowledgements}
Data collection and sharing for this project was funded by the Alzheimer's Disease Neuroimaging Initiative (ADNI) (National Institutes of Health Grant U01 AG024904) and 
DOD ADNI (Department of Defense award number W81XWH-12-2-0012). ADNI is funded by the National Institute on Aging, the National Institute of Biomedical Imaging and Bioengineering, and through generous contributions from the following: AbbVie, Alzheimer's Association; Alzheimer's Drug Discovery Foundation; Araclon Biotech; BioClinica, Inc.; Biogen; Bristol-Myers Squibb Company; CereSpir, Inc.; Cogstate; Eisai Inc.; Elan Pharmaceuticals, Inc.; Eli Lilly and Company; EuroImmun; F. Hoffmann-La Roche Ltd and its affiliated company Genentech, Inc.; Fujirebio; GE Healthcare; IXICO Ltd.; Janssen Alzheimer Immunotherapy Research \& Development, LLC.; Johnson \& Johnson Pharmaceutical Research \& Development LLC.; Lumosity; Lundbeck; Merck \& Co., Inc.; Meso Scale Diagnostics, LLC.; NeuroRx Research; Neurotrack Technologies; Novartis Pharmaceuticals Corporation; Pfizer Inc.; Piramal Imaging; Servier; Takeda Pharmaceutical Company; and Transition Therapeutics. The Canadian Institutes of Health Research is providing funds to support ADNI clinical sites in Canada. Private sector contributions are facilitated by the Foundation for the National Institutes of Health (www.fnih.org). The grantee organization is the Northern California Institute for Research and Education, and the study is coordinated by the Alzheimer's Therapeutic Research Institute at the University of Southern California. ADNI data are disseminated by the Laboratory for Neuro Imaging at the University of Southern California.

\bibliography{ref}

\begin{thebibliography}{32}
\providecommand{\natexlab}[1]{#1}
\providecommand{\url}[1]{\texttt{#1}}
\expandafter\ifx\csname urlstyle\endcsname\relax
  \providecommand{\doi}[1]{doi: #1}\else
  \providecommand{\doi}{doi: \begingroup \urlstyle{rm}\Url}\fi

\bibitem[Abbasi et~al.(2023)Abbasi, Deng, Magsi, Ali, Kumar, and Zubedi]{abbasi2023optimizing}
Erum~Yousef Abbasi, Zhongliang Deng, Arif~Hussain Magsi, Qasim Ali, Kamlesh Kumar, and Asma Zubedi.
\newblock Optimizing skin cancer survival prediction with ensemble techniques.
\newblock \emph{Bioengineering}, 11\penalty0 (1):\penalty0 43, 2023.

\bibitem[Albert et~al.(2011)Albert, DeKosky, and Dickson]{Albert2011ADJ}
M.~S. Albert, S.~T. DeKosky, and D.~et~al. Dickson.
\newblock The diagnosis of mild cognitive impairment due to alzheimer’s disease: Recommendations from the national institute on aging–alzheimer’s association workgroups.
\newblock \emph{Alzheimer’s \& Dementia}, 7\penalty0 (3):\penalty0 270--279, 2011.
\newblock \doi{10.1016/j.jalz.2011.03.008}.

\bibitem[Association(2024)]{2024ADJ_facts_figures}
Alzheimer's Association.
\newblock 2024 alzheimer's disease facts and figures.
\newblock \emph{Alzheimer's \& Dementia}, 20\penalty0 (3):\penalty0 487--588, 2024.
\newblock \doi{10.1002/alz.13700}.
\newblock URL \url{https://alz-journals.onlinelibrary.wiley.com/doi/10.1002/alz.13700}.

\bibitem[Blacker et~al.(2007)Blacker, Lee, Muzikansky, Martin, Tanzi, McArdle, Moss, and Albert]{blacker2007neuropsychological}
Deborah Blacker, Hang Lee, Alona Muzikansky, Emily~C Martin, Rudolph Tanzi, John~J McArdle, Mark Moss, and Marilyn Albert.
\newblock Neuropsychological measures in normal individuals that predict subsequent cognitive decline.
\newblock \emph{Archives of neurology}, 64\penalty0 (6):\penalty0 862--871, 2007.

\bibitem[Blennow et~al.(2010)Blennow, Hampel, Weiner, and Zetterberg]{blennow2010cerebrospinal}
Kaj Blennow, Harald Hampel, Michael Weiner, and Henrik Zetterberg.
\newblock Cerebrospinal fluid and plasma biomarkers in alzheimer disease.
\newblock \emph{Nature Reviews Neurology}, 6\penalty0 (3):\penalty0 131--144, 2010.

\bibitem[Chen and Guestrin(2016)]{chen2016xgboost}
Tianqi Chen and Carlos Guestrin.
\newblock Xgboost: A scalable tree boosting system.
\newblock In \emph{Proceedings of the 22nd acm sigkdd international conference on knowledge discovery and data mining}, pages 785--794, 2016.

\bibitem[Chen et~al.(2017)Chen, Denny, Harvey, Farias, Mungas, DeCarli, and Beckett]{chen2017progression}
Yingjia Chen, Katherine~G Denny, Danielle Harvey, Sarah~Tomaszewski Farias, Dan Mungas, Charles DeCarli, and Laurel Beckett.
\newblock Progression from normal cognition to mild cognitive impairment in a diverse clinic-based and community-based elderly cohort.
\newblock \emph{Alzheimer's \& Dementia}, 13\penalty0 (4):\penalty0 399--405, 2017.

\bibitem[Cui et~al.(2011)Cui, Liu, Luo, Zhen, Fan, Liu, Zhu, Park, Jiang, Jin, et~al.]{cui2011identification}
Yue Cui, Bing Liu, Suhuai Luo, Xiantong Zhen, Ming Fan, Tao Liu, Wanlin Zhu, Mira Park, Tianzi Jiang, Jesse~S Jin, et~al.
\newblock Identification of conversion from mild cognitive impairment to alzheimer's disease using multivariate predictors.
\newblock \emph{PloS one}, 6\penalty0 (7):\penalty0 e21896, 2011.

\bibitem[Ghali et~al.(2020)Ghali, Usman, Chellube, Degm, Hoti, Umar, and Abba]{ghali2020advanced}
UM~Ghali, Abdullahi~Garba Usman, ZM~Chellube, Mohamed Alhosen~Ali Degm, Kujtesa Hoti, Huzaifah Umar, and SI~Abba.
\newblock Advanced chromatographic technique for performance simulation of anti-alzheimer agent: An ensemble machine learning approach.
\newblock \emph{SN Applied Sciences}, 2:\penalty0 1--12, 2020.

\bibitem[Harrell et~al.(1982)Harrell, Califf, Pryor, Lee, and Rosati]{harrell1982evaluating}
Frank~E Harrell, Robert~M Califf, David~B Pryor, Kerry~L Lee, and Robert~A Rosati.
\newblock Evaluating the yield of medical tests.
\newblock \emph{Jama}, 247\penalty0 (18):\penalty0 2543--2546, 1982.

\bibitem[Heagerty et~al.(2000)Heagerty, Lumley, and Pepe]{heagerty2000time}
Patrick~J Heagerty, Thomas Lumley, and Margaret~S Pepe.
\newblock Time-dependent roc curves for censored survival data and a diagnostic marker.
\newblock \emph{Biometrics}, 56\penalty0 (2):\penalty0 337--344, 2000.

\bibitem[Hothorn et~al.(2006)Hothorn, B{\"u}hlmann, Dudoit, Molinaro, and Van Der~Laan]{hothorn2006survival}
Torsten Hothorn, Peter B{\"u}hlmann, Sandrine Dudoit, Annette Molinaro, and Mark~J Van Der~Laan.
\newblock Survival ensembles.
\newblock \emph{Biostatistics}, 7\penalty0 (3):\penalty0 355--373, 2006.

\bibitem[Ishwaran et~al.(2008)Ishwaran, Kogalur, Blackstone, and Lauer]{ishwaran2008random}
Hemant Ishwaran, Udaya~B Kogalur, Eugene~H Blackstone, and Michael~S Lauer.
\newblock Random survival forests.
\newblock \emph{The Annals of Applied Statistics}, pages 841--860, 2008.

\bibitem[Jack et~al.(2013)Jack, Knopman, and Jagust]{Jack2013Lancet-Neurology}
C.~R. Jack, D.~S. Knopman, and W.~J. et~al. Jagust.
\newblock Tracking pathophysiological processes in alzheimer’s disease: An updated hypothetical model of dynamic biomarkers.
\newblock \emph{The Lancet Neurology}, 12\penalty0 (2):\penalty0 207--216, 2013.
\newblock \doi{10.1016/S1474-4422(12)70291-0}.

\bibitem[Jack et~al.(2010)Jack, Knopman, Jagust, Shaw, Aisen, Weiner, Petersen, and Trojanowski]{jack2010hypothetical}
Clifford~R Jack, David~S Knopman, William~J Jagust, Leslie~M Shaw, Paul~S Aisen, Michael~W Weiner, Ronald~C Petersen, and John~Q Trojanowski.
\newblock Hypothetical model of dynamic biomarkers of the alzheimer's pathological cascade.
\newblock \emph{The Lancet Neurology}, 9\penalty0 (1):\penalty0 119--128, 2010.

\bibitem[Javeed et~al.(2022)Javeed, Dallora, Berglund, and Anderberg]{javeed2022intelligent}
Ashir Javeed, Ana~Luiza Dallora, Johan~Sanmartin Berglund, and Peter Anderberg.
\newblock An intelligent learning system for unbiased prediction of dementia based on autoencoder and adaboost ensemble learning.
\newblock \emph{Life}, 12\penalty0 (7):\penalty0 1097, 2022.

\bibitem[Katzman et~al.(2018)Katzman, Shaham, Cloninger, Bates, Jiang, and Kluger]{katzman2018deepsurv}
Jared~L Katzman, Uri Shaham, Alexander Cloninger, Jonathan Bates, Tingting Jiang, and Yuval Kluger.
\newblock Deepsurv: personalized treatment recommender system using a cox proportional hazards deep neural network.
\newblock \emph{BMC medical research methodology}, 18:\penalty0 1--12, 2018.

\bibitem[Li and Luo(2019)]{li2019dynamic}
Kan Li and Sheng Luo.
\newblock Dynamic predictions in bayesian functional joint models for longitudinal and time-to-event data: An application to alzheimer’s disease.
\newblock \emph{Statistical methods in medical research}, 28\penalty0 (2):\penalty0 327--342, 2019.

\bibitem[Li et~al.(2017)Li, Chan, Doody, Quinn, Luo, Initiative, et~al.]{Li2017JAD}
Kan Li, Wenyaw Chan, Rachelle~S Doody, Joseph Quinn, Sheng Luo, Alzheimer’s Disease~Neuroimaging Initiative, et~al.
\newblock Prediction of conversion to alzheimer’s disease with longitudinal measures and time-to-event data.
\newblock \emph{Journal of Alzheimer's Disease}, 58\penalty0 (2):\penalty0 361--371, 2017.

\bibitem[Li et~al.(2018)Li, O'Brien, Lutz, Luo, Initiative, et~al.]{Li2018ADJ}
Kan Li, Richard O'Brien, Michael Lutz, Sheng Luo, Alzheimer's Disease~Neuroimaging Initiative, et~al.
\newblock A prognostic model of alzheimer's disease relying on multiple longitudinal measures and time-to-event data.
\newblock \emph{Alzheimer's \& Dementia}, 14\penalty0 (5):\penalty0 644--651, 2018.

\bibitem[McKhann et~al.(1984)McKhann, Drachman, Folstein, Katzman, Price, and Stadlan]{McKhann1984Neurology}
Guy McKhann, David Drachman, Marshall Folstein, Robert Katzman, Donald Price, and Emanuel~M Stadlan.
\newblock Clinical diagnosis of alzheimer's disease: Report of the nincds-adrda work group under the auspices of department of health and human services task force on alzheimer's disease.
\newblock \emph{Neurology}, 34\penalty0 (7):\penalty0 939--939, 1984.

\bibitem[Minka(2000)]{minka2000bayesian}
Thomas Minka.
\newblock Bayesian linear regression.
\newblock Technical report, Citeseer, 2000.

\bibitem[Parent et~al.(2023)Parent, Rousseau, Predovan, Duchesne, and Hudon]{Parent2023Aging-Brain}
C.~Parent, L.~S. Rousseau, D.~Predovan, S.~Duchesne, and C.~Hudon.
\newblock Longitudinal association between $\beta$-amyloid accumulation and cognitive decline in cognitively healthy older adults: A systematic review.
\newblock \emph{Aging Brain}, 3:\penalty0 100074, 2023.
\newblock \doi{10.1016/j.nbas.2023.100074}.

\bibitem[Petersen et~al.(1999)Petersen, Smith, Waring, Ivnik, Tangalos, and Kokmen]{Petersen1999AN}
Ronald~C Petersen, Glenn~E Smith, Stephen~C Waring, Robert~J Ivnik, Eric~G Tangalos, and Emre Kokmen.
\newblock Mild cognitive impairment: clinical characterization and outcome.
\newblock \emph{Archives of neurology}, 56\penalty0 (3):\penalty0 303--308, 1999.

\bibitem[P{\"o}lsterl et~al.(2016)P{\"o}lsterl, Gupta, Wang, Conjeti, Katouzian, and Navab]{polsterl2016heterogeneous}
Sebastian P{\"o}lsterl, Pankaj Gupta, Lichao Wang, Sailesh Conjeti, Amin Katouzian, and Nassir Navab.
\newblock Heterogeneous ensembles for predicting survival of metastatic, castrate-resistant prostate cancer patients.
\newblock \emph{F1000Research}, 5, 2016.

\bibitem[Sperling et~al.(2014)Sperling, Mormino, and Johnson]{sperling2014evolution}
Reisa Sperling, Elizabeth Mormino, and Keith Johnson.
\newblock The evolution of preclinical alzheimer’s disease: implications for prevention trials.
\newblock \emph{Neuron}, 84\penalty0 (3):\penalty0 608--622, 2014.

\bibitem[Syed et~al.(2020)Syed, Khan, Hassan, Alromema, Binsawad, and Alsayed]{syed2020ensemble}
Asif~Hassan Syed, Tabrej Khan, Atif Hassan, Nashwan~A Alromema, Muhammad Binsawad, and Alhuseen~Omar Alsayed.
\newblock An ensemble-learning based application to predict the earlier stages of alzheimer’s disease (ad).
\newblock \emph{IEEE Access}, 8:\penalty0 222126--222143, 2020.

\bibitem[Tibshirani(1997)]{tibshirani1997lasso}
Robert Tibshirani.
\newblock The lasso method for variable selection in the cox model.
\newblock \emph{Statistics in medicine}, 16\penalty0 (4):\penalty0 385--395, 1997.

\bibitem[Wang et~al.(2018)Wang, Lu, Zheng, Su, Zhou, Chen, and Zhang]{wang2018early}
Bing Wang, Kun Lu, Xiao Zheng, Benyue Su, Yuming Zhou, Peng Chen, and Jun Zhang.
\newblock Early stage identification of alzheimer's disease using a two-stage ensemble classifier.
\newblock \emph{Current Bioinformatics}, 13\penalty0 (5):\penalty0 529--535, 2018.

\bibitem[White et~al.(2011)White, Royston, and Wood]{White2011SIM_MICE}
Ian~R. White, Patrick Royston, and Angela~M. Wood.
\newblock Multiple imputation using chained equations: issues and guidance for practice.
\newblock \emph{Statistics in Medicine}, 30\penalty0 (4):\penalty0 377--399, February 2011.
\newblock \doi{10.1002/sim.4067}.
\newblock URL \url{https://doi.org/10.1002/sim.4067}.

\bibitem[Wu(2012)]{wu2012elastic}
Yichao Wu.
\newblock Elastic net for cox’s proportional hazards model with a solution path algorithm.
\newblock \emph{Statistica Sinica}, 22:\penalty0 27, 2012.

\bibitem[Yao et~al.(2022)Yao, Frydman, Larocque, and Simonoff]{yao2022ensemble}
Weichi Yao, Halina Frydman, Denis Larocque, and Jeffrey~S Simonoff.
\newblock Ensemble methods for survival function estimation with time-varying covariates.
\newblock \emph{Statistical Methods in Medical Research}, 31\penalty0 (11):\penalty0 2217--2236, 2022.

\end{thebibliography}

\section*{Appendix}
\appendix

\section{Survival Models in the Ensemble Framework}
\label{appendix}

The survival models described below, including Random Survival Forest (RSF), DeepSurv, and XGBoost, form the ensemble framework for predicting preclinical Alzheimer's disease progression, as detailed in Section~\ref{sec:methods}. Their risk scores contribute to performance evaluation using time-dependent Area Under the Curve (AUC) and Integrated AUC (iAUC), computed as described in Section~\ref{sec:methods-evaluation} at 100 evenly spaced time points via inverse probability of censoring weighting (IPCW) and numerical integration (trapezoidal rule) over the observed survival range.

\subsection{Random Survival Forest (RSF)}

The Random Survival Forest (RSF) \cite{ishwaran2008random} is a nonparametric method for survival analysis, extending the random forest framework to model relationships between predictors and survival times. RSF builds an ensemble of survival trees, each constructed from a bootstrap sample of the data, capturing approximately $63.2\%$ of unique observations due to sampling with replacement. For each tree, the algorithm recursively splits the data into homogeneous subsets based on predictor variables. At each node, a random subset of predictors is selected, and the optimal variable and split point are chosen to maximize survival differences between child nodes, typically using the log-rank test statistic. Splitting halts when a predefined criterion is met, such as a minimum number of events per node or maximum tree depth. Each terminal node (leaf) contains survival information for its observations.

Mathematically, RSF estimates the cumulative hazard function at each terminal node using the Nelson-Aalen estimator:
$$
\hat{H}(t \mid X)=\sum_{i: t_i \leqslant t} \frac{\delta_i}{R\left(t_i\right)},
$$
where $t_i$ is the event or censoring time for the $i$-th individual, $\delta_i$ is an indicator (1 if the event occurs, 0 if censored), and $R\left(t_i\right)$ is the number of individuals at risk at time $t_i$. The survival function is then:
$$
\hat{S}(t \mid X)=\exp (-\hat{H}(t \mid X)) .
$$

Ensemble estimates are obtained by averaging across all $B$ trees in the forest:
$$
\hat{H}_{\mathrm{RSF}}(t \mid X)=\frac{1}{B} \sum_{b=1}^B \hat{H}_b(t \mid X), \quad \hat{S}_{\mathrm{RSF}}(t \mid X)=\exp \left(-\hat{H}_{\mathrm{RSF}}(t \mid X)\right),
$$
where $\hat{H}_b(t \mid X)$ is the cumulative hazard function from the $b$-th tree.
RSF excels in handling high-dimensional datasets and capturing complex, nonlinear interactions without assuming a specific hazard function. This flexibility makes it well-suited for medical research, such as analyzing survival data with multiple predictors to identify key factors in Alzheimer's disease progression, as shown in this study.

\subsection{DeepSurv (DS)}

DeepSurv \cite{katzman2018deepsurv} is a deep learning method for survival analysis that uses neural networks to model complex relationships between covariates and survival times. It defines the hazard function as $h(t \mid \boldsymbol{X})=h_0(t) \exp (f(\boldsymbol{X}))$, where $h_0(t)$ is the baseline hazard and $f(\boldsymbol{X})$ is a neural network function of covariates $\boldsymbol{X}$. Implemented as a fully connected feedforward neural network, it employs multiple layers with nonlinear activation functions like ReLU (Rectified Linear Unit) or Tanh. ReLU is preferred for its simplicity and mitigation of vanishing gradient issues, while Tanh provides a bounded output range suitable for specific cases.

Dropout layers, which randomly disable a fraction of units during training, prevent overfitting and promote robust feature learning. The network's depth (number of layers) and width (neurons per layer) can be tuned to balance complexity and computational needs, with deeper designs capturing intricate patterns but requiring more data. The output $f(\boldsymbol{X})$ serves as the log-risk function, driving the partial likelihood:
$$
L(\theta)=\prod_{i=1}^n\left(\frac{\exp \left(f_\theta\left(\boldsymbol{X}_i\right)\right)}{\sum_{j \in R\left(t_i\right)} \exp \left(f_\theta\left(\boldsymbol{X}_j\right)\right)}\right)^{\delta_i},
$$
where $\theta$ denotes the network parameters, $\delta_i$ is the event indicator (1 if observed, 0 if censored), and $R\left(t_i\right)$ is the risk set at time $t_i$. Training minimizes the negative log-partial likelihood:
$$
\mathcal{L}(\theta)=-\sum_{i=1}^n \delta_i\left(f_\theta\left(\boldsymbol{X}_i\right)-\log \sum_{j \in R\left(t_i\right)} \exp \left(f_\theta\left(\boldsymbol{X}_j\right)\right)\right),
$$
using optimizers such as Stochastic Gradient Descent (SGD) or Adam to adjust weights and learn the hazard function.

\begin{figure}[htbp]
    \centering
    \includegraphics[width=0.5\textwidth]{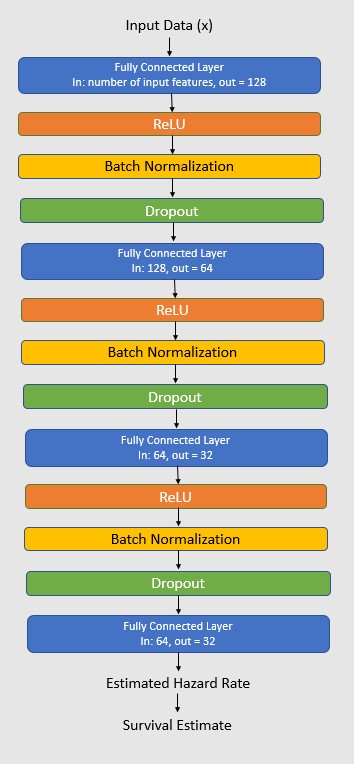}
    \caption{Network architecture of DeepSurv used for survival prediction in Alzheimer's disease progression.}
    \label{fig:deepsurv}
\end{figure}

In this study, DeepSurv exploits longitudinal biomarker trajectories to predict cognitive decline in Alzheimer's disease, with its architecture tailored to high-dimensional data (see Figure~\ref{fig:deepsurv}). This adaptability enhances its capacity to detect nonlinear dynamics essential for early risk stratification.

\subsection{XGBoost (XGB)}

XGBoost (Extreme Gradient Boosting) \cite{chen2016xgboost} is an efficient gradient boosting method for survival analysis, designed for high performance across machine learning tasks. It constructs an ensemble by sequentially adding decision trees, each correcting the residuals of prior trees to optimize a loss function. For survival analysis, XGBoost adapts the Cox proportional hazards model to manage right-censored data, modeling the risk score as $\eta_i=x_i^{\top} \beta$ for individual $i$, where $x_i$ is the covariate vector and $\beta$ represents learned weights.
The risk score is iteratively refined at each step $m$ :
$$
\eta_i^{(m)}=\eta_i^{(m-1)}+\alpha f_m\left(x_i\right),
$$
where $\alpha$ is the learning rate and $f_m\left(x_i\right)$ is the $m$-th tree's output. Optimization minimizes a Cox partial likelihood-derived loss using first- and second-order gradients. The gradient for individual $i$ is:
$$
g_i=\delta_i-\frac{\exp \left(\eta_i\right)}{\sum_{j \in R\left(t_i\right)} \exp \left(\eta_j\right)},
$$
indicating error correction direction, while the Hessian is:
$$
h_i=\frac{\exp \left(\eta_i\right) \sum_{j \in R\left(t_i\right)} \exp \left(\eta_j\right)-\exp \left(\eta_i\right)^2}{\left(\sum_{j \in R\left(t_i\right)} \exp \left(\eta_j\right)\right)^2},
$$
reflecting loss curvature to refine step sizes. Here, $\delta_i$ is the event indicator (1 if observed, 0 if censored), $t_i$ is the event or censoring time, and $R\left(t_i\right)$ is the risk set at $t_i$. These gradients guide tree construction via local loss approximation.

To curb overfitting, XGBoost includes a regularization term, $\Omega\left(f_m\right)=\gamma T+\frac{1}{2} \lambda \sum_{j=1}^T w_j^2$, where $T$ is the number of leaves, $w_j$ is the leaf weight, $\gamma$ controls tree complexity, and $\lambda$ applies $L_2$ regularization. The objective function at each iteration is:
$$
\mathrm{Obj}^{(m)}=\sum_{i=1}^n\left[g_i f_m\left(x_i\right)+\frac{1}{2} h_i f_m\left(x_i\right)^2\right]+\Omega\left(f_m\right) .
$$

This balances accuracy and simplicity, improving generalization.
XGBoost offers advanced features like parallel processing, weighted quantile sketches for sparse data, and missing value handling, making it scalable to large, high-dimensional datasets as used here. Additional tools—cross-validation, early stopping, and feature importance ranking—support tuning and interpretation. In this study, XGBoost harnesses longitudinal biomarkers to predict Alzheimer's disease progression, providing robustness to multicollinearity and versatility in modeling complex survival patterns, complementing the ensemble approach.

\section*{Supplementary Analysis: Age-Related Heterogeneity in Predictive Performance}

This supplementary analysis examines age-related heterogeneity in predictive performance for preclinical Alzheimer’s disease (AD) progression from cognitively normal (CN) to mild cognitive impairment (MCI) or AD over the full follow-up period, using the ensemble survival framework described in the main manuscript (Sections 3 and 4). The analysis utilizes longitudinal biomarker data from the Alzheimer’s Disease Neuroimaging Initiative (ADNI). Of the 721 participants reported in the main text, 15 were excluded due to ages outside the 60–90 range, resulting in 706 participants grouped into three age groups: 61–70 (N=229), 71–80 (N=383), and 81–90 (N=94). Performance was assessed across three visit settings: baseline, a 6-month follow-up (Visit 1), and a 12-month follow-up (Visit 2), using Harrell’s concordance index (C-index) and time-dependent area under the curve (AUC) as discrimination metrics.

Table~\ref{tab:age-performance} presents performance metrics for four ensemble survival models, Cox LASSO with Ensemble Averaging (EA), Cox LASSO with Bayesian Model Averaging (BMA), Cox Elastic Net with EA, and Cox Elastic Net with BMA, across the age groups and visit settings. A consistent trend shows younger individuals (61–70) achieving the highest C-index and AUC values across all settings, while older individuals (81–90) exhibit lower predictive accuracy, reflecting greater heterogeneity in older age groups. Subgroup gains differ from the overall cohort average (e.g., 48.2\% AUC gain in the main text, Section 5.1).

At baseline, using only initial data, predictive performance is modest, with C-index values ranging from 0.567 to 0.630 and AUC from 0.483 to 0.626 across groups. The 61–70 group performs best (e.g., C-index 0.630 for Cox Elastic + BMA), while the 81–90 group shows the lowest accuracy (e.g., AUC 0.483 for Cox Elastic + BMA). Incorporating longitudinal data from the 6-month follow-up significantly enhances performance, with C-index values reaching 0.851–0.915 and AUC values 0.846–0.937. For the 61–70 group, this represents a 45.2\% C-index increase from 0.630 to 0.915 and a 50.2\% AUC gain from 0.624 to 0.937; for 71–80, a 39.5\% C-index increase from 0.623 to 0.869 and a 43.0\% AUC gain from 0.616 to 0.881; and for 81–90, a 50.1\% C-index increase from 0.567 to 0.851 and a 75.2\% AUC gain from 0.483 to 0.846, demonstrating the value of repeated measurements. The 61–70 group peaks at 0.915 (C-index, Cox Elastic + BMA), while the 81–90 group lags at 0.851. The 12-month follow-up yields the highest performance, with C-index values reaching 0.859–0.931 and AUC values 0.855–0.955, though age-related disparities persist. For the 61–70 group, this reflects an additional 1.8\% C-index gain from 0.915 to 0.931 and a 1.9\% AUC gain from 0.937 to 0.955; for 71–80, a 2.5\% C-index gain from 0.869 to 0.891 and a -0.2\% AUC gain from 0.881 to 0.879 (indicating no improvement); and for 81–90, a 0.9\% C-index gain from 0.851 to 0.859 and a 1.1\% AUC gain from 0.846 to 0.855, indicating diminishing returns, as noted in Section 4.2.

Bayesian Model Averaging (BMA) consistently outperforms Ensemble Averaging (EA) across visit settings, especially in multi-visit scenarios. For instance, in the three-visit setting, Cox Elastic + BMA achieves a C-index of 0.931 for the 61–70 group, compared to 0.929 for Cox Elastic + EA, suggesting BMA’s incorporation of model uncertainty enhances accuracy when leveraging longitudinal data.

These findings highlight that younger individuals benefit more from longitudinal data, likely due to more predictable progression patterns, while older individuals face challenges from heterogeneity. The smaller sample size in the 81–90 group (N=94) may limit stability, and ADNI’s homogeneity restricts generalizability.

\begin{table}[h!]
\centering
\caption{Predictive Performance by Age Group and Visit Setting}
\label{tab:age-performance}
\resizebox{\textwidth}{!}{%
\begin{tabular}{@{}lcccccc@{}}
\toprule
\textbf{Setting / Model} & \multicolumn{3}{c}{\textbf{C-Index (95\% CI)}} & \multicolumn{3}{c}{\textbf{Time-Dependent AUC (95\% CI)}} \\ 
\cmidrule(lr){2-4} \cmidrule(lr){5-7}
 & \textbf{61–70 (N=229)} & \textbf{71–80 (N=383)} & \textbf{81–90 (N=94)} & \textbf{61–70} & \textbf{71–80} & \textbf{81–90} \\ \midrule
\textbf{Baseline} & & & & & & \\
Cox LASSO + EA & 0.627 (0.607, 0.647) & 0.619 (0.599, 0.639) & 0.566 (0.519, 0.613) & 0.626 (0.600, 0.652) & 0.617 (0.594, 0.641) & 0.485 (0.431, 0.539) \\
Cox LASSO + BMA & 0.628 (0.609, 0.646) & 0.619 (0.599, 0.639) & 0.568 (0.519, 0.618) & 0.626 (0.601, 0.651) & 0.617 (0.594, 0.641) & 0.486 (0.428, 0.547) \\
Cox Elastic + EA & 0.630 (0.610, 0.650) & 0.622 (0.607, 0.636) & 0.564 (0.508, 0.620) & 0.625 (0.602, 0.647) & 0.616 (0.592, 0.639) & 0.481 (0.420, 0.541) \\
Cox Elastic + BMA & 0.630 (0.609, 0.651) & 0.623 (0.608, 0.638) & 0.567 (0.517, 0.616) & 0.624 (0.602, 0.647) & 0.616 (0.592, 0.640) & 0.483 (0.431, 0.536) \\ \midrule
\textbf{6-Month Follow-Up (Two Visits)} & & & & & & \\
Cox LASSO + EA & 0.902 (0.892, 0.912) & 0.859 (0.831, 0.886) & 0.867 (0.855, 0.879) & 0.925 (0.914, 0.935) & 0.881 (0.842, 0.919) & 0.859 (0.834, 0.884) \\
Cox LASSO + BMA & 0.904 (0.895, 0.913) & 0.862 (0.838, 0.886) & 0.869 (0.856, 0.881) & 0.927 (0.918, 0.935) & 0.883 (0.845, 0.922) & 0.859 (0.835, 0.883) \\
Cox Elastic + EA & 0.912 (0.892, 0.932) & 0.856 (0.812, 0.900) & 0.868 (0.860, 0.876) & 0.934 (0.914, 0.954) & 0.878 (0.833, 0.924) & 0.847 (0.828, 0.865) \\
Cox Elastic + BMA & 0.915 (0.896, 0.934) & 0.869 (0.862, 0.878) & 0.851 (0.814, 0.902) & 0.937 (0.919, 0.955) & 0.881 (0.834, 0.928) & 0.846 (0.827, 0.865) \\ \midrule
\textbf{12-Month Follow-Up (Three Visits)} & & & & & & \\
Cox LASSO + EA & 0.926 (0.911, 0.941) & 0.895 (0.880, 0.911) & 0.847 (0.825, 0.869) & 0.948 (0.936, 0.960) & 0.879 (0.860, 0.897) & 0.811 (0.772, 0.849) \\
Cox LASSO + BMA & 0.928 (0.914, 0.941) & 0.898 (0.883, 0.912) & 0.849 (0.828, 0.869) & 0.949 (0.938, 0.960) & 0.880 (0.861, 0.898) & 0.809 (0.774, 0.844) \\
Cox Elastic + EA & 0.929 (0.918, 0.941) & 0.888 (0.868, 0.908) & 0.857 (0.830, 0.885) & 0.954 (0.945, 0.962) & 0.878 (0.863, 0.894) & 0.856 (0.820, 0.891) \\
Cox Elastic + BMA & 0.931 (0.919, 0.943) & 0.891 (0.875, 0.907) & 0.859 (0.837, 0.883) & 0.955 (0.946, 0.964) & 0.879 (0.867, 0.892) & 0.855 (0.819, 0.891) \\ \bottomrule
\end{tabular}%
}
\begin{flushleft}
\footnotesize{Metrics are means across 5-fold cross-validation, with 95\% confidence intervals derived from Rubin’s Rules for 20 imputed datasets.}
\end{flushleft}
\end{table}

\end{document}